\documentclass[runningheads]{llncs}
\usepackage{cite}
\usepackage{amsmath,amssymb,amsfonts}
\usepackage{graphicx}
\usepackage{textcomp}
\usepackage{xcolor}
\usepackage{tabularx}
\usepackage{array}
\usepackage{booktabs}
\usepackage{multirow}
\usepackage{hyperref}
\usepackage{pgfplots}
\usetikzlibrary{positioning}
\usepackage{caption}
\usepackage{subcaption}

\usepackage{comment}

\makeatletter
\newcommand{\removelatexerror}{\let\@latex@error\@gobble}
\makeatother

\usepackage[ruled,vlined]{algorithm2e}

\begin{document}
\title{Towards Robust Deep Neural Networks for Affect and Depression Recognition from Speech}
\author{Alice Othmani\inst{1}\orcidID{0000-0002-3442-0578} \and
Daoud Kadoch\inst{2} \and
Kamil Bentounes\inst{2} \and
Emna Rejaibi\inst{3} \and
Romain Alfred\inst{4} \and 
Abdenour Hadid\inst{5}}
\authorrunning{A. Othmani et al.}
\institute{University of Paris-Est Créteil, Vitry sur Seine, France
\email{alice.othmani@u-pec.fr}\\ \and
Sorbonne University, Paris, France \and
INSAT, Tunis, Tunisie \and
ENSIIE, Évry, France \and
Polytechnic University of Hauts-de-France, Valenciennes, France 
}
\maketitle              % typeset the header of the contribution
\begin{abstract}
Intelligent monitoring systems and affective computing applications have emerged in recent years to enhance healthcare. Examples of these applications include assessment of affective states such as Major Depressive Disorder (MDD). MDD describes the constant expression of certain emotions: negative emotions (low Valence) and lack of interest (low Arousal). High-performing intelligent systems would enhance MDD diagnosis in its early stages. In this paper, we present a new deep neural network architecture, called EmoAudioNet, for emotion and depression recognition from speech. Deep EmoAudioNet learns from the time-frequency representation of the audio signal and the visual representation of its spectrum of frequencies. Our model shows very promising results in predicting affect and depression. It works similarly or outperforms the state-of-the-art methods according to several evaluation metrics on RECOLA and on DAIC-WOZ datasets in predicting arousal, valence, and depression. Code of EmoAudioNet is publicly available on GitHub: \url{https://github.com/AliceOTHMANI/EmoAudioNet}

\end{abstract}
\section{Introduction}

Artificial Emotional Intelligence (EI) or affective computing has attracted increasing attention from the scientific community. Affective computing consists of endowing machines with the ability to recognize, interpret, process and simulate human affects. Giving machines skills of emotional intelligence is an important key to enhance healthcare and further boost the medical assessment of several mental disorders.

Affect describes the experience of a human's emotion resulting from an interaction with stimuli. Humans express an affect through facial, vocal, or gestural behaviors. A happy or angry person will typically speak louder and faster, with strong frequencies, while a sad or bored person will speak slower with low frequencies. Emotional arousal and valence are the two main dimensional affects used to describe emotions. Valence describes the level of pleasantness, while arousal describes the intensity of excitement. A final method for measuring a user’s affective state is to ask questions and to identify emotions during an interaction. Several post-interaction questionnaires exist for measuring affective states like the Patient Health Questionnaire 9 (PHQ-9) for depression recognition and assessment. The PHQ is a self report questionnaire of nine clinical questions where a score ranging from 0 to 23 is assigned to describe Major Depressive Disorder (MDD) severity level. MDD is a mental disease which affects more than 300 million people in the world \cite{i1}, \textit{i.e.}, 3\% of the worldwide population. The psychiatric taxonomy classifies MDD among the low moods \cite{i3}, \textit{i.e.}, a condition characterised by a tiredness and a global physical, intellectual, social and emotional slow-down. In this way, the speech of depressive subjects is slowed, the pauses between two speakings are lengthened and the tone of the voice (prosody) is more monotonous. 

In this paper, a new deep neural networks architecture, called EmoAudioNet, is proposed and evaluated for real-life affect and depression recognition from speech. The remainder of this article is organised as follows. Section~\ref{related_work} introduces related works with affect and depression recognition from speech. Section~\ref{sec:motivations} introduces the motivations behind this work. Section~\ref{proposed_approach} describes the details of the overall proposed method. Section~\ref{experiments_and_results} describes the entire experiments and the extensive experimental results. Finally, the conclusion and future work are presented in Section~\ref{conclusion}.
\section{Related work}
\label{related_work}
Several approaches are reported in the literature for affect and depression recognition from speech. These methods can be generally categorized into two groups: hand-crafted features-based approaches and deep learning-based approaches.

\subsection{Handcrafted features-based approaches}

In this family of approaches, there are two main steps~: feature extraction and classification. An overview of handcrafted features-based approaches for affect and depression assessment from speech is presented in Table~\ref{overview_hand_crafted_depression_assessment}.

\subsubsection{Handcrafted features}
Acoustic Low-Level Descriptors (LDD) are extracted from the audio signal. These LLD are grouped into four main categories: the \textbf{spectral LLD} (Harmonic Model and Phase Distortion Mean (HMPDM0-24), etc.), the \textbf{cepstral LLD} (Mel-Frequency Cepstral Coefficients (MFCC) \cite{ i10, i24}, etc.), the \textbf{prosodic LLD} (Formants \cite{f1}, etc.), and the \textbf{voice quality LLD} (Jitter, and Shimmer \cite{t1}, etc.). A set of statistical features are also calculated (max, min, variance and standard deviation of LLD \cite{i11, i23}). Low \textit{et al.} \cite{t5} propose the experimentation of the Teager Energy Operator (TEO) based features. 

A comparison of the performances of the prosodic, spectral, glottal (voice quality), and TEO features for depression recognition is realized in \cite{t5} and it demonstrates that the different features have similar accuracies. The fusion of the prosodic LLD and the glottal LLD based models seems to not significantly improve the results, or decreased them. However, the addition of the TEO features improves the performances up to +31,35\% for depressive male. 

\subsubsection{Classification of Handcrafted features}

Comparative analysis of the performances of several classifiers in depression assessment and prediction indicate that the use of an hybrid classifier using Gaussian Mixture Models (GMM) and Support Vector Machines (SVM) model gave the best overall classification results\cite{i14, t5}. Different fusion methods, namely feature, score and decision fusion have been also investigated in \cite{i14} and it has been demonstrated that : first, amongst the fusion methods, score  fusion  performed  better  when  combined  with  GMM, HFS and MLP classifiers. Second, decision fusion worked best for SVM (both for raw data and GMM models) and finally, feature fusion  exhibited  weak  performance compared to other fusion methods. 

\subsection{Deep learning-based approaches}

Recently, approaches based on deep learning have been proposed \cite{i18, ddl2, ddl3, ddl4, ddl5, ddl6, ddl7, ddl8, ddl10}. Several handcrafted features are extracted from the audio signals and fed to the deep neural networks, except in Jain \cite{ddl6} where only the MFCC are considered. In other approaches, raw audio signals are fed to deep neural networks \cite{tt8}. An overview of deep learning-based methods for affect and depression assessment from speech is presented in Table~\ref{overview_deep_depression_assessment}.

Several deep neural networks have been proposed. Some deep architectures are based on feed-forward neural networks \cite{i22, tt9, ddl3}, some others are based on convolutional neural networks such as \cite{ddl6} and \cite{i18} whereas some others are based on recurrent neural networks such as \cite{i24} and \cite{ddl2}.
A comparative study \cite{ddl4} of some neural networks, BLSTM-MIL, BLSTM-RNN, BLSTM-CNN, CNN, DNN-MIL and DNN, demonstrates that the BLSTM-MIL outperforms the other studied architectures.
Whereas, in Jain \cite{ddl6}, the Capsule Network is demonstrated as the most efficient architecture, compared to the BLSTM with Attention mechanism, CNN and LSTM-RNN. For the assessment of the level of depression using the Patient Health Questionnaire 8 (PHQ-8), Yang \textit{et al.} \cite{i18} exerts a DCNN. To the best of our knowledge, their approach outperforms all the existing approaches on DAIC-WOZ dataset.

%Several deep neural networks have been proposed : feed-forward neural network (FF-NN) \cite{i22, tt9, ddl3}, convolutional neural network (CNN) \cite{ddl6}, deep convolutional neural network (DCNN) \cite{i18}, deconvolutional neural network (DNN) \cite{ddl8, ddl10}, long short-term memory recurrent neural network (LSTM-RNN) \cite{i22, i24, tt9, ddl2, ddl6}, bidirectional long short-term memory recurrent neural network (BLSTM-RNN) \cite{i22, tt9, ddl4}, deep bidirectional long short-term memory recurrent neural network (DBLSTM-RNN) \cite{i23}, LSTM-CNN \cite{tt8}, BLSTM-CNN \cite{ddl4}, DCNN-DNN \cite{ddl5}, BLSTM with Attention mechanism \cite{ddl6}, deconvolutional neural network multiple instance learning (DNN-MIL) \cite{ddl4}, BLSTM-MIL \cite{ddl4} and Capsule Network \cite{ddl6}. \\%

%\begin{longtable}
\begin{table*}[]
\centering
\caption{Overview of Shallow Learning based methods for Affect and Depression Assessment from Speech. (*) Results obtained over a group of Females.}
\begin{center}
\begin{tabularx}{16cm}{p{1.5cm}|X|X|p{2cm}|p{2.5cm}|p{2cm}}
%{17cm}{p{2cm}|X|X|p{1.5cm}|p{2.5cm}|p{2cm}}
\hline
Ref & Features & Classification & Dataset & Metrics & Value 
     \\ \hline
     Valstar \textit{et al.} \cite{i10} & prosodic + voice & SVM + grid search + & DAIC-WOZ & F1-score & 0.410 (0.582)\\
      & quality + spectral & random forest & & Precision & 0.267 (0.941)\\
      & & & & Recall & 0.889 (0.421)\\
      & & & & RMSE (MAE) & 7.78 (5.72)\\
      \hline
      Dhall \textit{et al.} \cite{t1} & energy + spectral + voicing quality + duration features & non-linear chi-square kernel & AFEW 5.0 & \textit{unavailable} & \textit{unavailable}\\
      \hline
      Ringeval \textit{et al.}  & prosodic LLD + voice  & random forest & SEWA & RMSE & 7.78\\
      \cite{i11} & quality + spectral & & & MAE & 5.72\\
      \hline
      Haq \textit{et al.} \cite{t3} & energy + prosodic + spectral + duration features & Sequential Forward Selection + Sequential Backward Selection + linear discriminant analysis +  Gaussian classifier uses Bayes decision theory & Natural speech databases & Accuracy & 66.5\%\\
      \hline
      Jiang \textit{et al.} \cite{i13} & MFCC + prosodic +  & ensemble logistic regression  & hand-crafted  & Males accuracy & 81.82\%(70.19\%*)\\
      & spectral LLD + glottal  & model for detecting  &dataset & Males sensitivity & 78.13\%(79.25\%*)\\
      & features & depression E algorithm & & Males specificity & 85.29\%(70.59\%*)\\
      \hline
      Low \textit{et al.} \cite{t5} & teager energy operator & Gaussian mixture model +  & hand-crafted  & Males accuracy& 86.64\%(78.87\%*)\\
      & based features &SVM & dataset& Males sensitivity & 80.83\%(80.64\%*)\\
      & & & & Males specificity& 92.45\%(77.27\%*)\\
      \hline
      Alghowinem \textit{et al.} \cite{i14} & energy + formants + glottal features + intensity + MFCC + prosodic + spectral + voice quality & Gaussian mixture model + SVM + decision fusion & hand-crafted dataset & Accuracy & 91.67\%\\
      \hline
      Valstar \textit{et al.} \cite{i16} & duration features+energy& correlation based feature  & AViD- & RMSE & 14.12\\
      &local min/max related functionals+spectral+voicing quality&selection + SVR + 5-flod cross-validation loop & Corpus& MAE & 10.35\\
      \hline
      Valstar \textit{et al.} \cite{t7} & duration features+energy& SVR & AVEC2014 & RMSE & 11.521\\
      &local min/max related functionals+spectral+voicing quality & & & MAE & 8.934\\
      \hline
      Cummins \textit{et al.} \cite{i20} & MFCC + prosodic + spectral centroid & SVM & AVEC2013 & Accuracy & 82\%\\
      \hline
      Lopez Otero \textit{et} & energy + MFCC +  & SVR & AVDLC & RMSE (MAE)& 8.88 (7,02)\\
      \textit{al.} \cite{i21} & prosodic + spectral\\
      \hline
      Meng \textit{et al.} \cite{t8} & spectral + energy & PLS regression & AVEC2013 & RMSE & 11.54\\
      & + MFCC + functionals & & & MAE & 9.78\\
      & features + duration features& & & CORR & 0.42\\

\hline
\end{tabularx}
\end{center}
\label{overview_hand_crafted_depression_assessment}
\end{table*}
%\end{longtable}

\begin{table*}[]

\caption{Overview of Deep Learning based methods for Affect and Depression Assessment from Speech. }
\begin{center}
\begin{tabularx}{16.5cm}{p{2cm}|p{4cm}|X|p{1.5cm}|p{4cm}|p{1.5cm}}
\hline
Ref & Features & Classification & Dataset & Metrics & Value \\
\hline
Yang \textit{et al.} \cite{i18} & spectral LLD + cepstral & DCNN & DAIC-WOZ & Depressed female RMSE & 4.590\\
&  LLD + prosodic LLD + & & & Depressed female MAE & 3.589\\
&voice quality LLD +  & & & Not depressed female RMSE & 2.864\\
&statistical functionals +   & & & Not depressed female MAE & 2.393\\
&regression functionals& & & Depressed male RMSE & 1.802\\
& & & & Depressed male MAE & 1.690\\
& & & & Not depressed male RMSE & 2.827\\
& & & & Not depressed male MAE & 2.575\\
\hline
Al Hanai \textit{et al.}  & spectral LLD + cepstral  & LSTM-RNN & DAIC & F1-score & 0.67\\
\cite{ddl2}& LLD + prosodic LLD + & & & Precision & 1.00\\
& voice quality LLD + & & & Recall & 0.50\\
& functionals& & & RMSE & 10.03\\
& & & & MAE & 7.60\\
\hline
Dham \textit{et al.} \cite{ddl3} & prosodic LLD + voice  & FF-NN & AVEC2016 & RMSE & 7.631\\
&quality LLD + functionals + BoTW & & & MAE & 6.2766\\
\hline
Salekin \textit{et al.} \cite{ddl4} & spectral LLD + MFCC +  & NN2Vec + BLSTM-MIL & DAIC-WOZ & F1-score & 0.8544\\
& functionals& & & Accuracy & 96.7\%\\
\hline
Yang \textit{et al.} \cite{ddl5} & spectral LLD + cepstral & DCNN-DNN & DAIC-WOZ & Female RMSE & 5.669\\
& LLD + prosodic LLD +  & & & Female MAE & 4.597\\
&voice quality LLD +  & & & Male RMSE & 5.590\\
& functionals& & & Male MAE & 5.107\\
\hline
Jain \cite{ddl6} & MFCC & Capsule Network & VCTK corpus & Accuracy & 0.925\\
\hline
Chao \textit{et al.} \cite{ddl7} & spectral LLD + cepstral LLD + prosodic LLD & LSTM-RNN & AVEC2014 & \textit{unavailable} & \textit{unavailable}\\
\hline
Gupta \textit{et al.} \cite{ddl8} & spectral LLD + cepstral LLD + prosodic LLD + voice quality LLD + functionals & DNN & AViD-Corpus & \textit{unavailable} & \textit{unavailable}\\
\hline
Kang \textit{et al.} \cite{ddl10} & spectral LLD + prosodic  & DNN & AVEC2014 & RMSE & 7.37\\
&LLD + articulatory features & SRI's submitted system to  & & MAE & 5.87\\
& & AVEC2014 & & Pearson's Product Moment  & 0.800\\
& & median-way score-level fusion & &Correlation coefficient & \\
\hline
Tzirakis \textit{et al.} \cite{tzirakis2018} & raw signal & CNN and 2-layers LSTM & RECOLA & loss function based on CCC & .440(arousal) \\
& & & & & .787(valence) \\

\hline
Tzirakis \textit{et al.} \cite{tt8} & raw signal & CNN and LSTM & RECOLA & CCC & .686(arousal) \\
 & & & & & .261(valence) \\
\hline
Tzirakis \textit{et al.} \cite{ddl1} & raw signal & CNN & RECOLA & CCC & .699(arousal) \\
& & & & & .311(valence) \\

% ----------------------------- DAOUD & KAMIL -----------------------------------

\hline
\end{tabularx}
\end{center}
\label{overview_deep_depression_assessment}
\end{table*}

\section{Motivations and Contributions}
\label{sec:motivations}
%A variety of features can be extracted from audio and speech for classification, as shown in Section~\ref{related_work}, either to represent the signal in the time domain or in the frequency domain. A good parametric representation of a signal is important to produce a better recognition performance.

Short-time spectral analysis is the most common way to characterize the speech signal using MFCCs. However, audio signals in their time-frequency representations, often present interesting patterns in the visual domain \cite{ddl11}. The visual representation of the spectrum of frequencies of a signal using its spectrogram shows a set of specific repetitive patterns. Surprisingly and to the best of our knowledge, it has not been reported in the literature a deep neural network architecture that combines information from time, frequency and visual domains for emotion recognition. 

The first contribution of this work is a new deep neural network architecture, called EmoAudioNet, that aggregate responses from a short-time spectral analysis and from time-frequency audio texture classification and that extract deep features representations in a learned embedding space. In a second contribution, we propose EmoAudioNet-based approach for instantaneous prediction of spontaneous and continuous emotions from speech. In particular, our specific contributions are as follows: (i) an automatic clinical depression recognition and assessment embedding network  (ii)  a small size two-stream CNNs to map audio data into two types of continuous emotional dimensions namely, arousal and valence and (iii) through experiments, it is shown that EmoAudioNet-based features outperforms the state-of-the art methods for predicting depression on DAIC-WOZ dataset and for predicting valence and arousal dimensions in terms of Pearson's Coefficient Correlation (PCC).
 
\begin{figure}[]
 \removelatexerror
 
\begin{algorithm}[H]

\SetAlgoLined
 \For{iteration in range(\(N\))}{
  \((\mathbf{X}_{\text{wav}}, \mathbf{y}_{\text{wav}})\) \(\leftarrow\) batch of input wav files and labels \newline
  \(\mathbf{e}_{\text{Spec}} \leftarrow f_{\Theta}(\mathbf{X}_{\text{wav}})\) Spectrogram features  \newline
  \(\mathbf{e}_{\text{MFCC}} \leftarrow f_{\phi}(\mathbf{X}_{\text{wav}})\) MFCC features \newline
  \(\mathbf{f}_{\text{MFCCSpec}} \leftarrow [\mathbf{e}_{\text{MFCC}},\mathbf{e}_{\text{Spec}}]\) Feature-level fusion \newline
  \(\mathbf{p}_{\text{MFCCSpec}} \leftarrow f_{\theta}(\mathbf{e}_{\text{MFCCSpec}})\) Predict class probabilities \newline
  \(L_\text{MFCCSpec} = \texttt{cross\_entropy\_loss}(\mathbf{p}_{\text{MFCCSpec}}, \mathbf{y}_\text{wav})\) \newline
  Obtain all gradients \(\Delta_\text{all} = (\frac{\partial L}{\partial \Theta}, \frac{\partial L}{\partial \phi})\) \newline
  \((\Theta, \phi, \theta) \leftarrow \texttt{ADAM}(\Delta_\text{all}\)) Update feature extractor and output heads' parameters simultaneously
 }
 \caption{EmoAudioNet embedding network. \newline Given two feature extractors \(f_{\Theta}\) and  \(f_{\phi}\),  number of training steps \(N\).}
\label{algo1}
\end{algorithm}

\end{figure}

\section{Proposed Method}
\label{proposed_approach}
\begin{figure*}[]
  \centering \includegraphics[width=\textwidth]{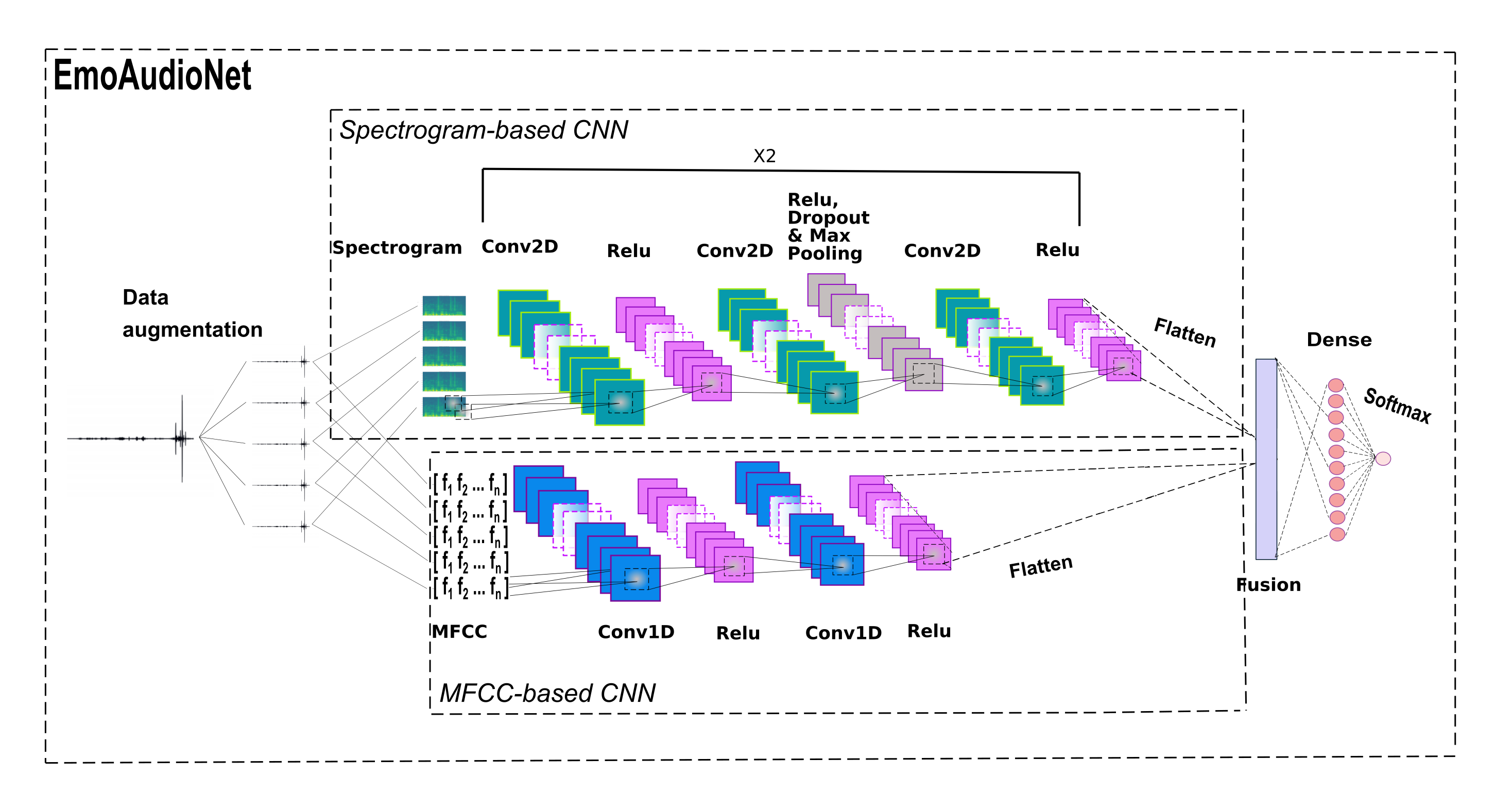}
  \caption{The diagram of the proposed deep neural networks architecture called EmoAudioNet. The output layer is dense layer of size n neurones with a Softmax activation function. n is defined according to the task. When the task concerns binary depression classification, n=2. When the task concerns depression severity level assessment, n=24. While, n=10 for arousal or valence prediction. }
  \label{diagram_proposed_approach}
\end{figure*}

We seek to learn a deep audio representation that is trainable end-to-end for emotion recognition. To achieve that, we propose a novel deep neural network called EmoAudioNet, which performs low-level and high-level features extraction and aggregation function learning jointly (See Algorithm.~\ref{algo1}). Thus, the input audio signal is fed to a small size two-stream CNNs that outputs the final classification scores. A data augmentation step is considered to increase the amount of data by adding slightly modified copies of already existing data.
The structure of EmoAudioNet presents three main parts as shown in Figure.~\ref{diagram_proposed_approach}~:
(i) An MFCC-based CNN, (ii) A spectrogram-based CNN and (iii) the aggregation of the responses of the MFCC-based and the spectrogram-based CNNs. In the following, more details about the three parts are given. 

\subsection{Data augmentation}
A data augmentation step is considered to overcome the problem of data scarcity by increasing the quantity of training data and also to improve the model's robustness to noise. Two different types of audio augmentation techniques are performed: \textbf{(1) Adding noise:} mix the audio signal with random noise. Each mix $z$ is generated using $z= x + \alpha \times rand(x)$ where $x$ is the audio signal and $\alpha$ is the noise factor. In our experiments, $\alpha=0.01$, $0.02$ and $0.03$. \textbf{(2) Pitch Shifting:} lower the pitch of the audio sample by 3 values (in semitones): (0.5, 2 and 5). 

\subsection{Spectrogram-based CNN stream}
%The spectrogram-based CNN aims to classify audio features in the visual domain. The texture-like time-frequency representation usually contains distinctive patterns that capture different characteristics. 
The spectrogram-based CNN presents low-level features descriptor followed by a high-level features descriptor. The Low-level features descriptor is the spectrogram of the input audio signal and it is computed as a sequence of Fast Fourier Transform (FFT) of windowed audio segments. The audio signal is split into 256 segments and the spectrum of each segment is computed. The Hamming window is applied to each segment. The spectrogram plot is a color image of $1900 \times 1200 \times 3$. The image is resized to $224 \times 224 \times 3$  before being fed to the High-level features descriptor. The high-Level features descriptor is a deep CNN, it takes as input the spectrogram of the audio signal. Its architecture, as shown in Fig.~\ref{diagram_proposed_approach}, is composed by two same blocks of layers. Each block is composed of a two-dimensional (2D) convolutional layer followed by a ReLU activation function, a second convolutional layer, a ReLU, a dropout and max pooling layer, a third convolutional layer and last ReLU.   

\subsection{MFCC-based CNN stream}
The MFCC-based CNN presents also a low-level followed by high-level features descriptors (see Fig.~\ref{diagram_proposed_approach}). The low-level features descriptor is the MFCC features of the input audio. To extract them, the speech signal is first divided into frames by applying a Hamming windowing function of 2.5s at fixed intervals of 500 ms. A cepstral feature vector is then generated and the Discrete Fourier Transform (DFT) is computed for each frame. Only the logarithm of the amplitude spectrum is retained. The spectrum is after smoothed and 24 spectral components into 44100 frequency bins are collected in the Mel frequency scale. The components of the Mel-spectral vectors calculated for each frame are highly correlated. Therefore, the Karhunen-Loeve (KL) transform is applied and is approximated by the Discrete Cosine Transform (DCT). Finally, 177 cepstral features are obtained for each frame. After the extraction of the MFCC features, they are fed to the high-Level features descriptor which is a small size CNN. To avoid overfitting problem, only two one-dimentional (1D) convolutional layers followed by a ReLU activation function each are performed.

\subsection{Aggregation of the spectrogram-based and MFCC-based responses}
Combining the responses of the two deep streams CNNs allows to study simultaneously the time-frequency representation and the texture-like time frequency representation of the audio signal. The output of the spectrogram-based CNN is a feature vector of size 1152, while the output of the MFCC-based CNN is a feature vector of size 2816. The responses of the two networks are concatenated and then fed to a fully connected layer in order to generate the label prediction of the emotion levels.

 \section{Experiments and results}
 \label{experiments_and_results}
 
 \subsection{Datasets}
Two publicly available datasets are used to evaluate the performances of EmoAudioNet:\\ 
\textbf{Dataset for affect recognition experiments}: RECOLA dataset \cite{ddl12} is a multimodal corpus of affective interactions in French. 46 subjects participated to data recordings. Only 23 audio recordings of 5 minutes of interaction are made publicly available and used in our experiments. Participants engaged in a remote discussion according to a survival task and six annotators measured emotion continuously on two dimensions: valence and arousal. \\ \textbf{Dataset for depression recognition and assessment experiments:} DAIC-WOZ depression dataset \cite{gratch2014distress} is introduced in the AVEC2017 challenge \cite{i11} and it provides audio recordings of clinical interviews of 189 participants. Each recording is labeled by the PHQ-8 score and the PHQ-8 binary. The PHQ-8 score defines the severity level of depression of the participant and the PHQ-8 binary defines whether the participant is depressed or not. For technical reasons, only 182 audio recordings are used. The average length of the recordings is 15 minutes with a fixed sampling rate of 16 kHz.

\subsection{Experimental Setup}
\textbf{Spectrogram-based CNN architecture: } The number of channels of the convolutional and pooling layers are both 128. While their filter size is $3 \times 3$. RELU is used as activation function for all the layers. The stride of the max pooling is 8. The dropout fraction is 0.1.  \\ 
 \textbf{MFCC-based CNN architecture: } The input is one-dimensional and of size $177 \times 1$. The filter size of its two convolutional layers is $5 \times 1$. RELU is used as activation function for all the layers. The dropout fraction is 0.1 and the stride of the max pooling is 8. \\ 
 \textbf{EmoAudioNet architecture: } The two features vectors are concatenated and fed to a fully connected layer of n neurones activated with a Softmax function. n is defined according to the task. When the task concerns binary depression classification, n=2. When the task concerns depression severity level assessment, n=24. While, n=10 for arousal or valence prediction.
 The ADAM optimizer is used. The learning rate is set experimentally to 10e-5 and it reduced when the loss value stops decreasing. The batch size is fixed to 100 samples. The number of epochs for training is set to 500. An early stopping is performed when the accuracy stops improving after 10 epochs. 
 % ---------------- RESULTATS ---------------------------

\begin{table*}[]
\caption{RECOLA dataset results for prediction
of arousal. The results obtained for the development and the test sets in term of three metrics: the accuracy, the Pearson's Coefficient Correlation (PCC) and the Root Mean Square error (RMSE).}
\begin{center}
    \setlength{\tabcolsep}{3pt}
    \begin{tabular}{|p{80pt}|p{40pt}|p{35pt}|p{35pt}|p{40pt}|p{35pt}|p{35pt}|}
            \hline
            & \multicolumn{3}{|c}{\textbf{Development}} & \multicolumn{3}{|c|}{\textbf{Test}} \\
 
             & Accuracy  & PCC  & RMSE  & Accuracy  & PCC  & RMSE \\
      \hline  \hline          
            MFCC-based CNN & 81.93\% & 0.8130 & 0.1501 & 70.23\% & 0.6981 & 0.2065 
            \\ \hline
            Spectrogram-based CNN & 80.20\% & 0.8157 & 0.1314 & 75.65\% & 0.7673 & 0.2099 
            \\ \hline
            \textbf{EmoAudioNet} & 94.49\% & 0.9521 & 0.0082 & 89.30\% & 0.9069 & 0.1229 
            \\ 
\hline           
\end{tabular}
 \end{center}
\label{results_affect_arousal}
\end{table*}

\begin{table*}[]
\caption{RECOLA dataset results for prediction
of valence. The results obtained for the development and the test sets in term of three metrics: the accuracy, the Pearson's Coefficient Correlation (PCC) and the Root Mean Square error (RMSE).}
\begin{center}
    \setlength{\tabcolsep}{3pt}
    \begin{tabular}{|p{80pt}|p{40pt}|p{35pt}|p{35pt}|p{40pt}|p{35pt}|p{35pt}|}
            \hline
            & \multicolumn{3}{|c}{\textbf{Development}} & \multicolumn{3}{|c|}{\textbf{Test}} \\
 
             & Accuracy  & PCC  & RMSE  & Accuracy  & PCC  & RMSE \\
      \hline  \hline          
            MFCC-based CNN & 83.37\% & 0.8289 & 0.1405 & 71.12\% & 0.6965 & 0.2082 
            \\ \hline
            Spectrogram-based CNN & 78.32\%  & 0.7984  & 0.1446  & 73.81\%  & 0.7598 & 0.2132
            \\ \hline
            \textbf{EmoAudioNet} & 95.42\%  & 0.9568 & 0.0625 & \textbf{91.44\%} & \textbf{0.9221} & \textbf{0.1118}
            \\ 
\hline           
\end{tabular}
 \end{center}
\label{results_affect_valence}
\end{table*}

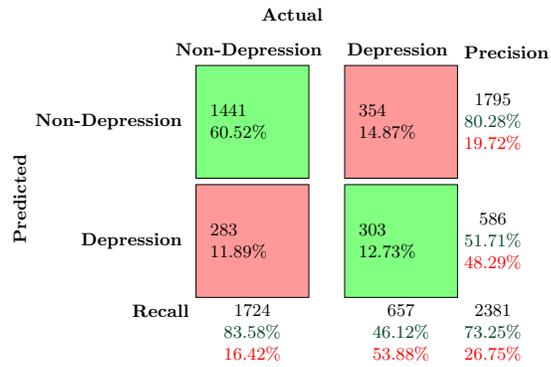
\begin{figure*}[]
		\centering \scalebox{.75}{\definecolor{forestgreen}{rgb}{0.0, 0.27, 0.13}

 \begin{tikzpicture}[align=center,
box/.style={draw,rectangle,minimum size=2cm,text width=1.5cm,align=left},
box_green/.style={draw,rectangle,fill=green!50,minimum size=2cm,text width=1.5cm,align=left},
box_red/.style={draw,rectangle,fill=red!40,minimum size=2cm,text width=1.5cm,align=left}]
\matrix (conmat) [row sep=.1cm,column sep=.1cm] {
    \node (box11) [box_green,
    label=left:\textbf{Non-Depression },
    label=above:\textbf{\textbf{Non-Depression }},
    ] {1441 \\ 60.52\%};
    &
    \node (box12) [box_red,
    label=above:\textbf{Depression },
    label=above right:\textbf{Precision},
    label=right: 1795\\ \textcolor{forestgreen}{80.28\%} \\ \textcolor{red}{19.72\%},
    ] {354 \\ 14.87\%};
    \\
    \node (box21) [box_red,
    label={left:\textbf{Depression }},
    label=below left:\textbf{Recall},
    label=below:1724\\ \textcolor{forestgreen}{83.58\%} \\ \textcolor{red}{16.42\%},
    ] {283 \\ 11.89\%};
    &
    \node (box22) [box_green,
    label=right:586\\ \textcolor{forestgreen}{51.71\%} \\ \textcolor{red}{48.29\%},
    label=below:657\\ \textcolor{forestgreen}{46.12\%} \\ \textcolor{red}{53.88\%},
    label=below right:2381\\ \textcolor{forestgreen}{73.25\%} \\ \textcolor{red}{26.75\%},
    ] {303 \\ 12.73\%};
    \\
};
 \node [rotate=90,left=.05cm of conmat,anchor=center,text width=1.5cm] {\textbf{Predicted}};
\node [above=.05cm of conmat] {\textbf{Actual}};
\end{tikzpicture}}
	\caption{Confusion Matrice of EmoAudioNet generated on the DAIC-WOZ test set}   
	\label{fig:confusionMatrix_depression}
\end{figure*}

\subsection{Experimental results on spontaneous and continuous emotion recognition from speech}

\subsubsection{Results of three proposed CNN architectures}
The experimental results of the three proposed architectures on predicting arousal and valence are given in Table~\ref{results_affect_arousal} and Table~\ref{results_affect_valence} . EmoAudioNet outperforms MFCC-based CNN and the spectrogram-based CNN with an accuracy of 89\% and 91\% for predicting aroural and valence respectively. The accuracy of the MFCC-based CNN is around 70\% and 71\% for arousal and valence respectively. The spectrogram-based CNN is slightly better than the MFCC-based CNN and its accuracy is 76\% for predicting arousal and 74\% for predicting valence.\\ 
EmoAudioNet has a Pearson Coefficient Correlation (PCC) of 0.91 for predicting arousal and 0.92 for predicting valence, and has also a Root Mean Square of Error (RMSE) of 0.12 for arousal's prediction and 0.11 for valence's prediction.\\

\subsubsection{Comparisons of EmoAudioNet and the stat-of-the art methods for arousal and valence prediction on RECOLA dataset} 
As shown in Table~\ref{comparison_results_RECOLA}, EmoAudioNet model has the best PCC of $0.9069$ for arousal prediction. In term of the RMSE, the approach proposed by He \textit{et al.} \cite{i23} outperforms all the existing methods with a RMSE equal to $0.099$ in predicting arousal. \\
For valence prediction, EmoAudioNet  outperforms state-of-the-art in predicting valence with a PCC of $0.9221$ without any fine-tuning. While the proposed approach by He \textit{et al.} \cite{i23} has the best RMSE of $0.104$.

\subsection{Experimental results on automatic clinical depression recognition and assessment}
EmoAudioNet framework is evaluated on two tasks on the DAIC-WOZ corpus. The first task is to predict depression from speech under the PHQ-8 binary. The second task is to predict the depression severity levels under the PHQ-8 scores.
\subsubsection{EmoAudioNet performances on depression recognition task}
EmoAudioNet is trained to predict the PHQ-8 binary (0 for non-depression and 1 for depression). The performances are summarized in Fig.~\ref{fig:confusionMatrix_depression}. The overall accuracy achieved in predicting depression reaches 73.25\% with an RMSE of 0.467. On the test set, 60.52\% of the samples are correctly labeled with non-depression, whereas, only 12.73\% are correctly diagnosed with depression. The low rate of correct classification of non-depression can be explained by the imbalance of the input data on the DAIC-WOZ dataset and the small amount of the participants labeled as depressed. F1 score is designed to deal with the non-uniform distribution of class labels by giving a weighted average of precision and recall. The non-depression F1 score reaches 82\% while the depression F1 score reaches 49\%. Almost half of the samples predicted with depression are correctly classified with a precision of 51.71\%. The number of non-depression samples is twice the number of samples labeled with depression. Thus, adding more samples of depressed participants would significantly increase the model's ability to recognize depression.

\subsubsection{EmoAudioNet performances on depression severity levels prediction task}
The depression severity levels are assessed by the PHQ-8 scores ranging from 0 for non-depression to 23 for severe depression. The RMSE achieved when predicting the PHQ-8 scores is 2.6 times better than the one achieved with the depression recognition task. The test loss reaches 0.18 compared to a 0.1 RMSE on the training set. 

\begin{table}[]
\caption{Comparisons of EmoAudioNet and the state-of-the art methods for arousal and valence prediction on RECOLA dataset. }
\begin{center}
    \begin{tabular}{l|c|c|c|c}
          %  \hline
          \hline
            & \multicolumn{2}{|c}{\textbf{Arousal}} & \multicolumn{2}{|c}{\textbf{Valence}}\\
 
            Method & PCC  & RMSE   & PCC  & RMSE \\
      %\hline            \hline 
      \hline
           He \textit{et al.} \cite{i23} & 0.836 & 0.099  & 0.529 & 0.104
           \\ 
           %\hline
           Ringeval \textit{et al.} \cite{i22} & 0.322 &  0.173 & 0.144 & 0.127
            \\ 
            %\hline
           EmoAudioNet & 0.9069 & 0.1229 & 0.9221 & 0.1118
            \\ \hline
%\hline
\end{tabular}
\end{center}
\label{comparison_results_RECOLA}
\end{table}

\begin{table}[]
\caption{Comparisons of EmoAudioNet and the stat-of-the art methods for prediction of depression on DAIC-WOZ dataset. (*) The results of the depression severity level prediction task. (**) for non-depression. ($\ddagger$) for depression.  ($^{Norm}$): Normalized RMSE  }
\begin{center}
    \begin{tabular}{l|c|c|r}
            \toprule
            Method & Accuracy  & RMSE & F1 Score\\
      \midrule            \midrule
             \multirow{2}{8em}{Yang \textit{et al.} \cite{i18}} & \textit{-} & 1.46 (*)  & \textit{-}\\
             & & (depressed male) &
    \\ \midrule
            \multirow{2}{8em}{Yang \textit{et al.} \cite{ddl5}} & \textit{-} & 5.59 (*) & \textit{-} \\
            & & (male) &
    \\ \midrule
            Valstar \textit{et al.} \cite{i10}& \textit{-}  & 7.78 (*) & \textit{-} 
    \\ \midrule
           Al Hanai \textit{et al.}\cite{ddl2}& \textit{-} & 10.03 & \textit{-}
    \\ \midrule
            Salekin \textit{et al.}\cite{ddl4}& 96.7\% & \textit{-} & 85.44\% 
    \\ \midrule
           \multirow{2}{8em}{Ma \textit{et al.}\cite{Depaudionet2016}} & \textit{-} & \textit{-} & 70\% (**)\\
           & & & 50\% ($\ddagger$)
    \\ \midrule
           \multirow{3}{8em}{Rejaibi \textit{et al.} \cite{MFCCbasedRNN2019}} & 76.27\% & 0.4 & 85\% (**)\\
           & & & 46\% ($\ddagger$) \\ 
           & - & $0.168^{Norm}$ (*) & -
    \\ \midrule
           \multirow{3}{8em}{EmoAudioNet} & 73.25\%  & 0.467 & 82\% (**) \\
           & & & 49\% ($\ddagger$) \\
           & - & $0.18^{Norm} |  4.14 $ (*) & -           
    \\  \bottomrule
    \end{tabular}
\end{center}
\label{comparison_results_DAIC-WOZ}
\end{table}

\subsubsection{Comparisons of EmoAudioNet and the state-of-the art methods for depression prediction on DAIC-WOZ dataset} 
Table~\ref{comparison_results_DAIC-WOZ} compares the performances of EmoAudioNet with the state-of-the-art approaches evaluated on the DAIC-WOZ dataset. To the best of our knowledge, in the literature, the best performing approach is the proposed approach in \cite{ddl4} with an F1 score of 85.44\% and an accuracy of 96.7\%. The proposed NN2Vec features with BLSTM-MIL classifier achieves this good performance thanks to the leave-one-speaker out cross-validation approach. Comparing to the other proposed approaches where a simple train-test split is performed, giving the model the opportunity to train on multiple train-test splits increase the model performances especially in small datasets. \\
In the depression recognition task, the EmoAudioNet outperforms the proposed architecture in \cite{Depaudionet2016} based on a Convolutional Neural Network followed by a Long Short-Term Memory network. The non-depression F1 score achieved with EmoAudioNet is better than the latter by 13\% with the exact same depression F1 score (50\%). \\
Moreover, the EmoAudioNet outperforms the LSTM network in \cite{MFCCbasedRNN2019} in correctly classifying samples of depression. The depression F1 score achieved with EmoAudioNet is higher than the MFCC-based RNN by 4\%. Meanwhile, the overall accuracy and loss achieved by the latter are better than EmoAudioNet by 2.14\% and 0.07 respectively.  According to the summarized results of previous works in Table~\ref{comparison_results_DAIC-WOZ}, the best results achieved so far in the depression severity level prediction task are obtained in \cite{MFCCbasedRNN2019}. The best normalized RMSE is achieved with the LSTM network to reach 0.168. EmoAudioNet reaches almost the same loss with a very low difference of 0.012. Our proposed architecture outperforms the rest of the results in the literature with the lowest normalized RMSE of $0.18$  in predicting depression severity levels (PHQ-8 scores) on the DAIC-WOZ dataset.

 \section{Conclusion and future work}
\label{conclusion}
In this paper, we proposed a new emotion and affect recognition methods from speech based on deep neural networks called EmoAudioNet. The proposed EmoAudioNet deep neural networks architecture is the aggregation of an MFCC-based CNN and a spectrogram-based CNN, which studies the time-frequency representation and the visual representation of the spectrum of frequencies of the audio signal. EmoAudioNet gives promising results and it approaches or outperforms state-of-art approaches of continuous dimensional affect recognition and automatic depression recognition from speech on RECOLA and DAIC-WOZ databases. In future work, we are planning (1) to improve the EmoAudioNet architecture with the given possible improvements in the discussion section and (2) to use EmoAudioNet architecture to develop a computer-assisted application for patient monitoring for mood disorders.


\begin{thebibliography}{8}

\bibitem{i1}
GBD 2015 Disease and Injury Incidence and Prevalence Collaborators: Global, regional, and national incidence, prevalence, and years lived with disability for 310 diseases and injuries, 1990-2015: a systematic analysis for the Global Burden of Disease Study 2015, \textit{Lancet, 388}, vol. 388, no 10053, pp. 1545-1602, 2015.

\bibitem{i3}
The National Institute of Mental Health: Depression,  \href{URL}{https://www.nimh.nih.gov/health/topics/depression/index.shtml}. Retrieved 2019, June 17.

%10
\bibitem{i10}
 Valstar, M., Gratch, J., Schuller, B.,  Ringeval, F., Lalanne, D., Torres Torres M.,  Scherer, S., Stratou, G., Cowie, R., Pantic, M. : Avec 2016 - Depression, mood, and emotion recognition workshop and challenge. In Proceedings of the 6th international workshop on audio/visual emotion challenge, pp. 3--10. ACM (2016).

\bibitem{i11}
Ringeval, F., Schuller, B., Valstar, V., Gratch, J., Cowie, R., Scherer, S., Mozgai, S., Cummins, N., Schmitt, M., Pantic, M. : Avec 2017 - Real-life depression, and affect recognition workshop and challenge. In Proceedings of the 7th Annual Workshop on Audio/Visual Emotion Challenge, pp. 3--9. ACM, (2017).

\bibitem{i13}
Jiang, H., Hu, B., Liu, Z., Wang, G., Zhang, L., Li, X., Kang, H. : Detecting Depression Using an Ensemble Logistic Regression Model Based on Multiple Speech Features. In Computational and mathematical methods in medicine, vol. 2018, (2018).

\bibitem{i14}
Alghowinem, S., Goecke, R., Wagner, M., Epps, J., Gedeon, T., Breakspear, M., Parker, G., : A comparative study of different classifiers for detecting depression from spontaneous speech. In 2013 IEEE International Conference on Acoustics, Speech and Signal Processing, pp. 8022--8026, (2013).

\bibitem{i16}
Valstar, M., Schuller, B., Smith, K., Eyben, F., Jiang, B., Bilakhia, S., Schnieder, S., Cowie, R., Pantic, M., : AVEC 2013: the continuous audio/visual emotion and depression recognition challenge. In Proceedings of the 3rd ACM international workshop on Audio/visual emotion challenge, pp. 3--10, (2013).

\bibitem{i18}
Yang, L., Sahli, H., Xia, X., Pei, E., Oveneke, M.C., Jiang, D., : Hybrid depression classification and estimation from audio video and text information. In Proceedings of the 7th Annual Workshop on Audio/Visual Emotion Challenge, pp. 45--51, ACM, (2017).

\bibitem{i20}
Cummins, N., Epps, J., Breakspear M., Goecke, R., : An investigation of depressed speech detection: Features and normalization.In Twelfth Annual Conference of the International Speech Communication Association, (2011). 

\bibitem{i21}
Lopez-Otero, P., Dacia-Fernandez, L., Garcia-Mateo, C., :  A study of acoustic features for depression detection. In 2nd International Workshop on Biometrics and Forensics, IEEE, pp. 1--6, (2014).

%20
\bibitem{i22}
Ringeval, F., Schuller, B., Valstar, M., Jaiswal, S., Marchi, E., Lalanne, D., Cowie R., Pantic, M., : Av+ ec 2015 - The first affect recognition challenge bridging across audio, video, and physiological data. In Proceedings of the 5th International Workshop on Audio/Visual Emotion Challenge, pp. 3--8, ACM, (2015).

\bibitem{i23}
He, L., Jiang, D., Yang, L., Pei, E., Wu, P., Sahli, H., : Multimodal affective dimension prediction using deep bidirectional long short-term memory recurrent neural networks. In Proceedings of the 5th International Workshop on Audio/Visual Emotion Challenge, pp. 73--80, ACM, 2015.

\bibitem{i24}
Ringeval, F., Schuller, B., Valstar, M., Cowie, R., Kaya, H., Schmitt, M., Amiriparian, S., Cummins, N., Lalanne, D., Michaud, A., Çiftçi, E., Güleç, H., Salah, A.A., Pantic, M., : AVEC 2018 workshop and challenge: Bipolar disorder and cross-cultural affect recognition. In Proceedings of the 2018 on Audio/Visual Emotion Challenge and Workshop, ACM, pp. 3--13, (2018).


\bibitem{t1}
Dhall, A., Ramana Murthy, O.V., Goecke, R., Joshi, J., Gedeon, T., : Video and image based emotion recognition challenges in the wild: Emotiw 2015. In Proceedings of the 2015 ACM on international conference on multimodal interaction, pp. 423--426, (2015).

%25
\bibitem{t3}
Haq, S., Jackson, P.J., Edge, J., :  Speaker-dependent audio-visual emotion recognition. In AVSP,  pp. 53--58, (2009).


\bibitem{t5}
Low, L.S.A.,  Maddage, N.C., Lech, M., Sheeber, L.B., Allen, N.B., :  Detection of clinical depression in adolescents' speech during family interactions. In IEEE Transactions on Biomedical Engineering, vol. 58, no. 3, pp. 574--586, (2010).

\bibitem{t7}
Valstar, M., Schuller, B.W., Krajewski, J., Cowie, R., Pantic, M., : AVEC 2014: the 4th international audio/visual emotion challenge and workshop. In Proceedings of the 22nd ACM international conference on Multimedia, pp. 1243--1244, (2014).

\bibitem{t8}
Meng, H., Huang, D., Wang, H., Yang, H., Ai-Shuraifi, M., Wang, Y., : Depression recognition based on dynamic facial and vocal expression features using partial least square regression. In Proceedings of the 3rd ACM international workshop on Audio/visual emotion challenge, pp. 21--30, (2013).


%%%Citations du table II
%30
\bibitem{tt8}
Trigeorgis, G., Ringeval, F., Brueckner, R., Marchi, E., Nicolaou, M.A., Schuller, B., Zafeiriou, S., : Adieu features? end-to-end speech emotion recognition using a deep convolutional recurrent network. In 2016 IEEE international conference on acoustics, speech and signal processing (ICASSP), pp. 5200--5204, (2016).

\bibitem{tt9}
Ringeval, F., Eyben, F., Kroupi, E., Yuce, A., Thiran, J.P., Ebrahimi, T., Lalanne, D., Schuller, B., : Prediction of asynchronous dimensional emotion ratings from audiovisual and physiological data. In Pattern Recognition Letters, vol. 66, pp. 22--30, (2015).

%%%Citations de la partie related work

%%%Citations de la partie hand-crafted features based

\bibitem{f1}
Ringeval, F., Schuller, B., Valstar, M., Cowie, R., Pantic, M., : AVEC 2015: The 5th international audio/visual emotion challenge and workshop. In Proceedings of the 23rd ACM international conference on Multimedia, pp. 1335--1336, (2015).

%review of deep learning tab

\bibitem{ddl1}
Tzirakis, P., Trigeorgis, G., Nicolaou, M.A., Schuller, B.W., Zafeiriou, S., : End-to-end multimodal emotion recognition using deep neural networks. In IEEE Journal of Selected Topics in Signal Processing, vol. 11, no. 8, pp. 1301--1309, (2017).

\bibitem{ddl2}
Al Hanai, T., Ghassemi, M.M., Glass, J.R., : Detecting Depression with Audio/Text Sequence Modeling of Interviews. In Interspeech, pp. 1716--1720, (2018).

\bibitem{ddl3}
Dham, S., Sharma, A., Dhall, A., : Depression scale recognition from audio, visual and text analysis. arXiv preprint arXiv:1709.05865.

\bibitem{ddl4}
Salekin, A., Eberle, J.W., Glenn, J.J., Teachman, B.A., Stankovic J.A., : A weakly supervised learning framework for detecting social anxiety and depression. In Proceedings of the ACM on interactive, mobile, wearable and ubiquitous technologies, vol. 2, no. 2, pp. 81, (2018).

\bibitem{ddl5}
Yang, L., Jiang, D., Xia, X., Pei, E., Oveneke, M.C., Sahli, H., : Multimodal measurement of depression using deep learning models. In Proceedings of the 7th Annual Workshop on Audio/Visual Emotion Challenge, pp. 53--59, (2017).

\bibitem{ddl6}
Jain, R., : Improving performance and inference on audio classification tasks using capsule networks. arXiv preprint arXiv:1902.05069, (2019).

\bibitem{ddl7}
Chao, L., Tao, J., Yang, M., Li, Y., : Multi task sequence learning for depression scale prediction from video. In 2015 International Conference on Affective Computing and Intelligent Interaction (ACII), IEEE, pp. 526--531, (2015).

\bibitem{ddl8}
Gupta, R., Sahu, S., Espy-Wilson, C.Y., Narayanan, S.S., : An Affect Prediction Approach Through Depression Severity Parameter Incorporation in Neural Networks. In Interspeech, pp. 3122--3126, (2017).


\bibitem{ddl10}
Kang, Y., Jiang, X., Yin, Y., Shang, Y., Zhou, X., : Deep transformation learning for depression diagnosis from facial images. In Chinese Conference on Biometric Recognition, Springer, Cham, pp. 13--22, (2017).

\bibitem{ddl11}
 Yu, G., and Slotine, J. J., : Audio classification from time-frequency texture. In 2009 IEEE International Conference on Acoustics, Speech and Signal Processing, pp. 1677--1680, (2009).

\bibitem{ddl12} 
Ringeval, F., Sonderegger, A., Sauer, J., Lalanne, D., : Introducing the RECOLA multimodal corpus of remote collaborative and affective interactions. In 2013 10th IEEE international conference and workshops on automatic face and gesture recognition (FG), IEEE, pp. 1--8, (2013).

\bibitem{gratch2014distress}
Gratch, J., Artstein, R., Lucas, G.M., Stratou, G., Scherer, S., Nazarian, A., Wood, R., Boberg, J., DeVault, D., Marsella, S., Traum, D.R., Rizzo, S., Morency, L.P., : The distress analysis interview corpus of human and computer interviews. LREC, pp. 3123--3128, (2014).


\bibitem{Depaudionet2016} Ma, X., Yang, H., Chen, Q., Huang, D., Wang, Y., : Depaudionet: An efficient deep model for audio based depression classification. In Proceedings of the 6th International Workshop on Audio/Visual Emotion Challenge, pp. 35--42, (2016).

\bibitem{MFCCbasedRNN2019} Rejaibi, E., Komaty, A., Meriaudeau, F., Agrebi, S., Othmani, A. : MFCC-based Recurrent Neural Network for Automatic Clinical Depression Recognition and Assessment from Speech. arXiv preprint arXiv:1909.07208, (2019).

\bibitem{tzirakis2018} Tzirakis, P.,  Zhang, J., Schuller, B.W. : End-to-end speech emotion recognition using deep neural networks. In 2018 IEEE International Conference on Acoustics, Speech and Signal Processing (ICASSP), pp. 5089--5093, (2018).

\bibitem{Poria2017}
Poria, S., Cambria, E., Bajpai, R., Hussain, A., : A review of affective computing: From unimodal analysis to multimodal fusion. In Information Fusion, vol. 37, pp. 98--125, (2017).

\bibitem{Rouast2019}
Rouast, P.V.,  Adam, M., Chiong, R., : Deep learning for human affect recognition: insights and new developments. In IEEE Transactions on Affective Computing, (2019).

\end{thebibliography}
\end{document}